\documentclass[reprint,amsmath,amssymb,aps,prl]{revtex4-1}

\usepackage{graphicx}
\usepackage{dcolumn}
\usepackage{bm}
\usepackage{hyperref}

\newcommand{\la}{\langle}
\newcommand{\ra}{\rangle}

\newcommand{\kF}{k_{\rm F}}

\newcommand{\be}{\begin{equation}}
\newcommand{\ee}{\end{equation}}

\begin{document}


\title{Momentum relaxation of a mobile impurity in a one-dimensional quantum gas.}

\author{E. Burovski$^1$}
\author{V. Cheianov$^1$}
\author{O. Gamayun$^{1,2}$}
\author{O. Lychkovskiy$^{1,3,4}$}
\affiliation{$^1$Physics Department, Lancaster University,  Lancaster LA1 4YB, United Kingdom}
\affiliation{$^2$Bogolyubov Institute for Theoretical Physics, 14-b Metrolohichna str., Kyiv 03680, Ukraine,}
\affiliation{$^3$Institute for Theoretical and Experimental Physics, 25 B. Cheremushkinskaya str.,
 Moscow 117218, Russia.}
\affiliation{$^4$ Russian Quantum Center, Novaya St. 100A, Skolkovo, Moscow Region, 143025, Russia.}

\date{\today}

\begin{abstract}
We investigate the time evolution of the momentum of an impurity atom injected
into a degenerate Tonks-Girardeau gas. We establish that given an initial momentum $p_0$
the impurity relaxes to a steady state with a non-vanishing momentum $p_\infty.$
The nature of the steady state is found to depend drastically on whether the masses of the impurity and the host are equal or not. This is due to multiple coherent scattering
processes leading to a resonant interaction between the impurity and the host in the case of equal masses.
The dependence of $p_\infty$ on $p_0$
remains non-trivial even in the limit of vanishing interaction between the impurity
and host particles. In this limit $p_\infty(p_0)$ is found explicitly.
\end{abstract}

\pacs{Valid PACS appear here}
\maketitle


{\em Introduction}.--- Laws governing the motion of a mobile particle through a fluid provide
a powerful insight into the fluid's dynamical properties. For this
reason impurities immersed in quantum fluids such as superfluid $^4$He and Fermi liquid
$^3$He have been a continued subject of study since late 1950's\cite{Khalat,*Meyer58,*prokof1993diffusion}. 
A remarkable class of quantum fluids, which are neither superfluids nor Fermi liquids
is found in one spatial dimension (1D) \cite{Giamarchi2003}.
Interest in impurities moving through such fluids started
with works on the Fermi edge singularity in inorganic quantum wires
\cite{Calleja1991,Ogawa1992,castella1993exact} and the mobility of a heavy
particle in a Luttinger liquid  \cite{CastroNeto1995}. Several distinctive
features of the motion of an impurity in 1D have been predicted theoretically
such as weak violation of superfluidity \cite{Astrakharchik2004},
non-Markovian relaxation patterns rooted in power-law singularities of the
fluid's spectral function \cite{imambekov2012one},
logarithmic subdiffusion \cite{Zvonarev2007}, quantum flutter
\cite{mathy2012quantum,knap2013quantum}
and quasi-Bloch momentum oscillations \cite{Gangardt2009}.

Early studies, both theoretical and experimental, were mainly concerned with
equilibrium spectral characteristics and linear response
properties of an impurity. At present, however, the focus is shifting
towards the analysis of dynamic, far-from-equilibrium phenomena \cite{Polkovnikov2011}.
This is fuelled by tremendous
experimental progress achieved over the past decade in the area of ultracold
atomic gases. First elongated traps containing interacting 1D Bose gases were
demonstrated ten years ago \cite{Moritz2003,Kinoshita2004,Paredes2004}.
By 2009 sufficient control over the system became available
to conduct experiments with an ensemble of impurity atoms
under a constant drag force \cite{Kohl2009} and out of
equilibrium impurity clouds injected into the
host \cite{catani2012quantum}.
Furthermore, development of single-atom-resolved control
\cite{spethmann2012dynamics} and
imaging techniques \cite{bakr2010probing,sherson2010single} opens unprecedented
experimental opportunities such as a direct observation
of the motion of an individual impurity atom in
a one-dimensional gas \cite{fukuhara2013quantum}.

On the theory side, several complementary approaches are being developed.
For bosonic hosts in the Bogoliubov limit remarkable results have been
obtained by methods of quantum hydrodynamics
\cite{Gangardt2009,Gangardt2012,schecter2012dynamics}.
In particular, it was predicted that the momentum of an impurity
driven by a constant force may exhibit oscillations resembling the Bloch
oscillations in an ideal crystal \cite{Gangardt2009} (however, this prediction was 
criticized in Ref. \cite{LL}).

A Bethe-Ansatz solvable model of an impurity injected in a
Tonks-Girardeau host has been considered in Ref. \cite{mathy2012quantum}.
Using numerical summation of form-factor series for a finite-size system
Ref. \cite{mathy2012quantum} investigated an impurity's momentum relaxation
at intermediate time scales.  The momentum of the impurity as a
function of time was found to follow a counterintuitive pattern
resembling underdamped periodic oscillations around some non-zero average.
This phenomenon was dubbed as the ``quantum flutter'' \cite{mathy2012quantum}.
Subsequent numerical simulations based on matrix product states extended
the results of Ref.\cite{mathy2012quantum} to a nonintegrable case
\cite{knap2013quantum}.
The results of Refs. \cite{mathy2012quantum,knap2013quantum} suggest, in particular,
the possibility of a non-vanishing steady-state momentum of the impurity.
Considering the absence of superfluidity in one
dimension \cite{Astrakharchik2004}, such incomplete momentum relaxation
contradicts equipartition of energy and signals the failure of
thermalization in the system. The purpose of the present Letter is
to explain the physical mechanism responsible for this phenomenon and
to develop a complete analytical theory of the formation of the
steady state in a certain perturbative limit.

To this end, we investigate the relaxation of the momentum of
an impurity weakly interacting with a degenerate ($T=0$)
Tonks-Girardeau gas \cite{Girardeau1960}, focusing on the infinite time steady state of the system.
Within the Boltzmann kinetic theory we find the dependence of the
infinite-time momentum of the impurity $p_\infty$ on the
initial momentum $p_0$, and explain
the mechanism by which $p_\infty$ is non-zero.
We find that when the masses of the impurity and the host particle are equal,
the Boltzmann theory breaks down.
In this case, we resort to an alternative approach based on the Bethe Ansatz
solution for a pointlike interaction. We first develop a novel method
of dealing with formfactor expansions numerically. Using the insight from
numerical simulations, we perform a controllable asymptotic analysis
of the problem to obtain a closed-form expression for $p_\infty(p_0)$.

{\em Problem formulation.}---We consider a single mobile impurity of mass $m_i$ immersed in the TG
gas of particles of mass $m_h.$  In the following we exploit the exact spectral equivalence
between the TG gas and a gas of non-interacting Fermi particles \cite{Girardeau1960}
and refer to the host particles as ``fermions.'' We assume a short-range repulsive
interaction between the impurity and the host fermions such that the
total Hamiltonian is
\be\label{Hamiltonian}
\hat H=\hat H_h^0+\hat H_i^0 +
\gamma \frac{k_F}{\pi m_h}\int dx \hat\rho_h(x)
\hat\rho_i(x).
\ee
Here $\hat H_h^0$, $\hat H_i^0$ and $\hat\rho_h(x)$, $\hat\rho_i(x)$ are the
Hamiltonians and density operators of the host fermions and the impurity,
respectively, and $\gamma$ is the dimensionless constant. Depending on the mass ratio $\eta \equiv m_i/m_h,$
we distinguish between the  cases of ``light impurity'',  $\eta<1$, and  ``heavy impurity'',
$\eta>1$. We define the Fermi momentum $\kF\equiv \pi \rho,$ where
$\rho$ is the host particle density.

We are interested in the time evolution of the system from an initial state
being a direct product of the ground state of $\hat H_h^0$  (which is merely a Fermi sea of the host particles), and a plane wave state of the impurity with momentum $p_0>0.$ Our main goal is to find the impurity momentum distribution function $w_p(t)$, investigate its  $t\to\infty$ limit, which we denote
by $w_{p_0\to p}^{\infty},$ and calculate the infinite-time momentum $p_\infty =  \sum_p p\,w_{p_0\to p}^{\infty}.$

\begin{figure}[t]
\begin{tabular}{cc}
\includegraphics[width=\linewidth]{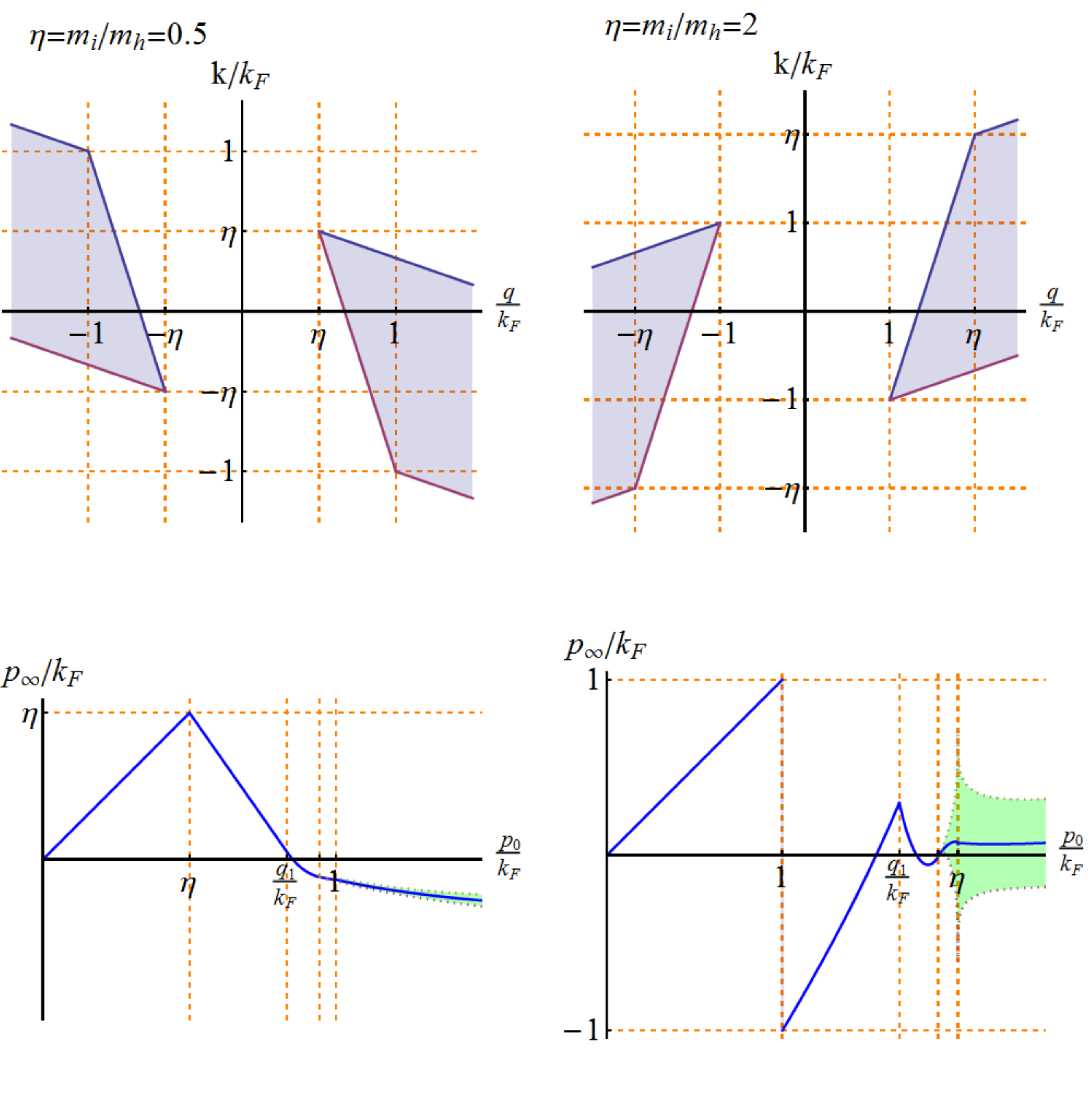}
\end{tabular}
\caption{
\label{fig momentum nonequal masses}
Upper panels: kinematically allowed regions for a single pairwise scattering in the cases of the light
(left) and heavy (right) impurity. The final momentum of the  impurity, $k$,   is shown vs the initial
momentum of the  impurity, $q.$ Lower panels: the impurity momentum at infinite time, $p_\infty,$
as a function of the initial momentum, $p_0$, for the light (left)
and heavy (right) impurity. Solid blue line shows an  iterative solution (two
iterations) of Eq. (\ref{integral eq for asymptotic momentum}).
Shaded area (green online) represents the maximum error: the exact
solution of Eq. (\ref{integral eq for asymptotic momentum}) lies inside this
area. Notice a much better convergence of iterations in the case of light
impurity.}
\end{figure}

{\em Kinematics.}---We begin our analysis from semiclassical considerations.
Kinematics of a two-particle scattering in 1D
is completely characterized by two momenta, {\it e.g.}
the initial, $q$, and final, $k$, momenta of the impurity.
Given these, the initial and final momenta of the host particle are
completely fixed
by two conservation laws.
In addition, the Pauli principe restricts
possible values of $k$ and  $q$  to a certain
region in the $(q,k)$ plane. This kinematically
allowed region is shown on Fig.~\ref{fig momentum nonequal
masses}. We see that the
allowed region exists for all $\vert q\vert>q_0,$ where
$q_0\equiv \kF \min\{1, \eta\},$ and its
boundaries are piecewise linear functions of
$q.$ We denote these functions by $u(q)$ for the upper boundary
and $d(q)$ for the lower one.

If the initial momentum of the impurity satisfies \mbox{$|p_0|<q_0,$}
Pauli blocking precludes any scattering, and the
impurity's momentum is conserved. For $|p_0|>q_0$, scattering events
continue until the impurity momentum drops below $q_0,$ after which
scattering stops. There exists a
momentum $q_1>q_0$ such that,  whenever $q_0<|p_0|<q_1$,
the impurity momentum drops below $q_0$ in a single scattering event.
Furthermore, there exists
an infinite ascending sequence $\{q_n\}$ such that $|p_0|<q_n$ implies
that the impurity momentum drops below $q_0$ in no more than $n$
collisions.
The recursive definition for the sequence reads
$ q_{n-1} =\max\{|u(q_n)|,|d(q_n)|\};$
the sequence converges to $q_\infty=\kF\max\{1,\eta \}$ with
$n\rightarrow\infty.$ Note that the case of equal masses ($\eta=1$) is special:
the whole sequence $\{q_n\}$ collapses to a single point, $q_n=\kF.$

Now consider the classical evolution of the momentum distribution function of an
impurity. If $|p_0|<q_n$ no more than $n$ collisions bring
the system to a steady state in which the impurity's momentum distribution
function $w_{p_0\to p}^{\infty}$ has a finite support $p\in [-q_0, q_0].$
There are no symmetries of the problem to prevent this function from having a
non-vanishing first moment. Therefore, in general $p_\infty\neq 0$. This conclusion is supported by a non-perturbative fully
quantum treatment: when $p_0$ lies in a certain range, one can rigorously prove that
$p_\infty$ is non-zero \cite{Lychkovskiy2013}.

{\em Applicability of Boltzmann equation.}---The above kinematical considerations are a good
starting point for the application of
Boltzmann's kinetic theory. The latter requires
the validity of Fermi golden rule, which is ensured by the smallness of the
dimensionless coupling constant, $\gamma\ll 1$, and by a narrow impurity level spacing (as compared to the collision rate) which implies $\gamma^2 N \gg 1$ for a system of $N$ particles.
Apart from these two constraints, the dimensionality of the problem imposes
extra conditions.
Indeed, the validity of Boltzmann's equation relies on the Lorentzian
shape of particle's spectral function such as in Fermi liquid theory
\cite{Mahan2000}. Generally in 1D systems
 spectral functions of particles exhibit essentially non-Lorentzian
shapes in the vicinity of the mass shell
\cite{Giamarchi2003,imambekov2012one,Zvonarev2009} (an expception from this rule was discussed in \cite{khodas2007fermi}).
The source of this peculiarity is virtual long wavelength modes which produce a logarithmically divergent contribution  to the self-energy,  $\delta \Sigma \sim \gamma^2 \ln N.$
In order for Boltzmann's equation to work one needs to suppress the divergence
and impose $\gamma^2\ln N \ll 1.$ Finally, another divergent contribution to self-energy arises from the ladder diagrams in the equal mass limit $m_i\to m_h.$ The physical meaning of this divergence will be explained below. The requirement that the ladder corrections can be neglected results in the condition
$
{|\eta-1|}\gg \gamma.
$

{\em Analysis of Boltzmann equation.}---
The kinetic equation reads as follows \cite{GG}:
\begin{gather}\label{Boltzmann equation}
\dot w_k(t)=-\Gamma_k w_k(t)+\sum_{q} \Gamma_{q
\rightarrow k} w_{q}(t),\\
\Gamma_{q\to k} =
\frac{\gamma^2}{\pi^2}\frac{\kF^2}{L m_h}\frac{\theta\bigl(d(q)<k<u(q)\bigr)}{
|q-k|}\,.
\label{Gammadef}
\end{gather}
Here $\Gamma_{q\to k}$ is the partial width and $\Gamma_k \equiv \sum_{q} \Gamma_{k \rightarrow q} $
is the total width. Due to kinematical constraints reflected in
step functions in Eq.~\eqref{Gammadef}, the
kinetic equation (\ref{Boltzmann equation})
leads to the following integral equation on the
asymptotic distribution:
\be\label{integral eq for asymptotic distr}
w_{p_0\rightarrow k}^\infty = \theta(q_0-|k|)\left( {\cal P}^{(1)}_{
p_0\rightarrow k}+\sum_{q\in {\cal R}(p_0)} {\cal P}^{(1)}_{ p_0\rightarrow q}
w_{q\rightarrow k}^\infty\right),
\ee
where  ${\cal P}^{(1)}_{ p_0\rightarrow k} = \Gamma_{p_0\to
k}/\Gamma_{p_0}$ is the probability that the impurity changes its momentum from
$p_0$ to $k$ in a {\it single} scattering event, and ${\cal R}(p_0)\equiv
[d(p_0),u(p_0)]\setminus[-q_0,q_0]$ is a kinematically determined integration
region.
Calculating the first moment of the distribution~\eqref{integral eq for asymptotic distr} (with respect to $k$) we find the
integral equation for the asymptotic
momentum,
\begin{gather}\label{integral eq for asymptotic momentum}
p_\infty (p_0) = p_\infty^{(1)} (p_0)
+\sum_{q \in{\cal R}(p_0)}  {\cal P}^{(1)}_{ p_0\rightarrow q} p_\infty
(q),\\ \nonumber
p_\infty^{(1)} (p_0)\equiv \sum_{k=-q_0}^{q_0} k {\cal P}^{(1)}_{
p_0\rightarrow k}.
\end{gather}
 For $|p_0|<q_0$ the momentum does not relax, $p_\infty
(p_0)=p_0.$
The coupling strength $\gamma$
cancels out from Eqs. \eqref{integral eq for asymptotic distr}, \eqref{integral eq for asymptotic momentum},
and in Eq. (\ref{Boltzmann equation}) it can be absorbed in rescaling of time.

Eqs. (\ref{Boltzmann equation}), (\ref{integral eq for
asymptotic distr}) and (\ref{integral eq for asymptotic momentum})
can be solved by iterations. In particular, the first iteration for asymptotic
momentum is $p_\infty^{(1)} (p_0)$. The $n$th iteration takes into account
classical evolution paths which involve no more than $n$ scattering events.
If $q_{n-1}<|p_0|<q_n,$ then $n$ iterations lead to an {\it exact}
solution. Fewer iterations give an approximate solution. If $|p_0| > q_\infty,$
any finite number of iterations gives an approximate solution. The convergence of
iterations is well controlled. The error at the $n$'th
step of iterative solution of Eq. (\ref{integral eq for asymptotic momentum})
is bounded from above by $q_0$ times the probability to scatter below
$q_0$ in more than $n$ collisions. The solution of Eq. (\ref{integral eq for
asymptotic momentum}) is plotted in Fig. \ref{fig momentum nonequal masses}. Note that since $p_{\infty}\in[-q_0,q_0]$,
the asymptotic velocity of an infinitely heavy impurity vanishes, which is consistent with Refs. \cite{Astrakharchik2004,PhysRevB.46.15233}.

The solution $p_\infty(p_0)$ of Eq.
(\ref{integral eq for asymptotic momentum})
has non-analyticities at $p_0=q_n,~n=0,1,...,\infty.$ The most
prominent one is a jump which occurs at $p_0=\kF$ when $m_i>m_h$, see Fig. \ref{fig momentum nonequal masses}.  In the vicinity of $\kF$ we find
\begin{equation}\label{asymptotic momentum heavy impurity}
p_{\infty}(p_0)=\left\{\begin{array}{ll}
p_0, & p_0<\kF \\
-\kF + \frac{m_i^2 + m_h^2 }{ m_i^2-m_h^2 }(p_0 - \kF),&p_0>\kF
\end{array}\right.
\end{equation}
Other type of singularity is a kink at $p_0=\kF\eta,$ where the velocity of the impurity equals the Fermi velocity,
resulting in a forward scattering anomaly.
We expect these singularities to be smoothed out by quantum
corrections in higher orders in $\gamma$.
Note that taking  the $m_i \rightarrow m_h$ limit
is non-trivial and is discussed below.

{\em Equal masses.}--- In this case, kinematics of a two-body collision reduces
to an exchange of momenta. Thus,
the very first scattering event brings the impurity to the state with
$k\in [-k_F, k_F]$, and creates a hole with momentum $-k$. From this moment on,
the impurity and a hole move with the same velocity. Further multiple coherent
scatterings in this two-body system allow the momentum of the impurity
to migrate unrestricted in the range $\left[-k_F, k_F\right]$. Thus one might expect
that $p_\infty = 0$. Below we demonstrate that this intuitive
expectation fails.

The Boltzmann equation does not capture multiple coherent scattering processes.
However, the contact interaction Hamiltonian \eqref{Hamiltonian} with
$m_i = m_h$ is Bethe Ansatz integrable
\cite{mcguire1965interacting,edwards1990magnetism,castella1993exact},
which allows us to calculate $p_{\infty}$ explicitly in the same limit,
$\gamma\to0$, $\gamma^2N\to\infty$, $\gamma^2\ln N\to 0$,
as for Boltzmann's theory above.
\footnote{Note that the breakdown of the Boltzmann kinetic equation
in the equal mass case is not directly related to integrability. In particular,
for a finite-range interaction potential the integrability is absent however
the Boltzmann equation fails for the same kinematical reasons.}


{\em Bethe Ansatz, form-factor expansion.}--- For a finite
number of particles of the background gas, $N$, eigenstates of the
Hamiltonian~\eqref{Hamiltonian}, $|\psi_\lambda\rangle$, are
labeled by ordered sets $\lambda=\{ n_0, n_1,\dots, n_N \}$ of $N+1$
distinct integers.
The value of the asymptotic momentum follows from the formfactor expansion,
\begin{equation}
p_\infty = \sum_{\lambda} \la \psi_\lambda |
\widehat{P_i} \, \bigl| \psi_\lambda \ra | \la \psi_\lambda| {\rm in} \ra \bigr|^2\;,
\label{ffactors}
\end{equation}
where $\widehat{P_i}$ is the impurity momentum operator, $ \vert {\rm in} \rangle$
is the initial state of the system, and the summation is over the infinite complete
set of eigenstates.
Explicit determinant
representations for the matrix elements entering \eqref{ffactors}
have been found in \cite{mathy2012quantum}. However, evaluation of the sum over intermediate states remains a challenge. A general analytical solution is unknown, and numerically the difficulty is to find an efficient and controllable way of selecting most relevant contributions to the Eq. \eqref{ffactors}.
While rather sophisticated ways of scanning the Hilbert space have been
developed \cite{CauxABACUS,mathy2012quantum}, the task remains difficult.
We have been able to both significantly advance numerical technique and develop successful analytic approach in the perturbative limit.

{\em Bethe Ansatz, numerics.}--- We note that the structure of Eq.\ \eqref{ffactors} naturally
lends itself to a \emph{stochastic} sampling of the Hilber space: instead of
evaluating the sum in Eq.\ \eqref{ffactors} in a pre-determined order, we
construct a random walk in the space of ordered sets $\lambda$,
based on the Metropolis algorithm \cite{Metropolis} with transition probabilities
proportional to $| \la
\psi_\lambda | \rm in \ra |^2$. This way, the algorithm automatically finds the most
relevant regions of the Hilbert space. In practice, we only use local updates of
the configurations (\textit{i.e.,} at each step of the Markov process
we only change one or two integers in the ordered set $\lambda$) and observe a very quick
convergence of the sum \eqref{ffactors}. Detailed description of the algorithm
will be given elsewhere \cite{Burovski2013}.%

\begin{figure}[h]
\includegraphics[width= \linewidth]{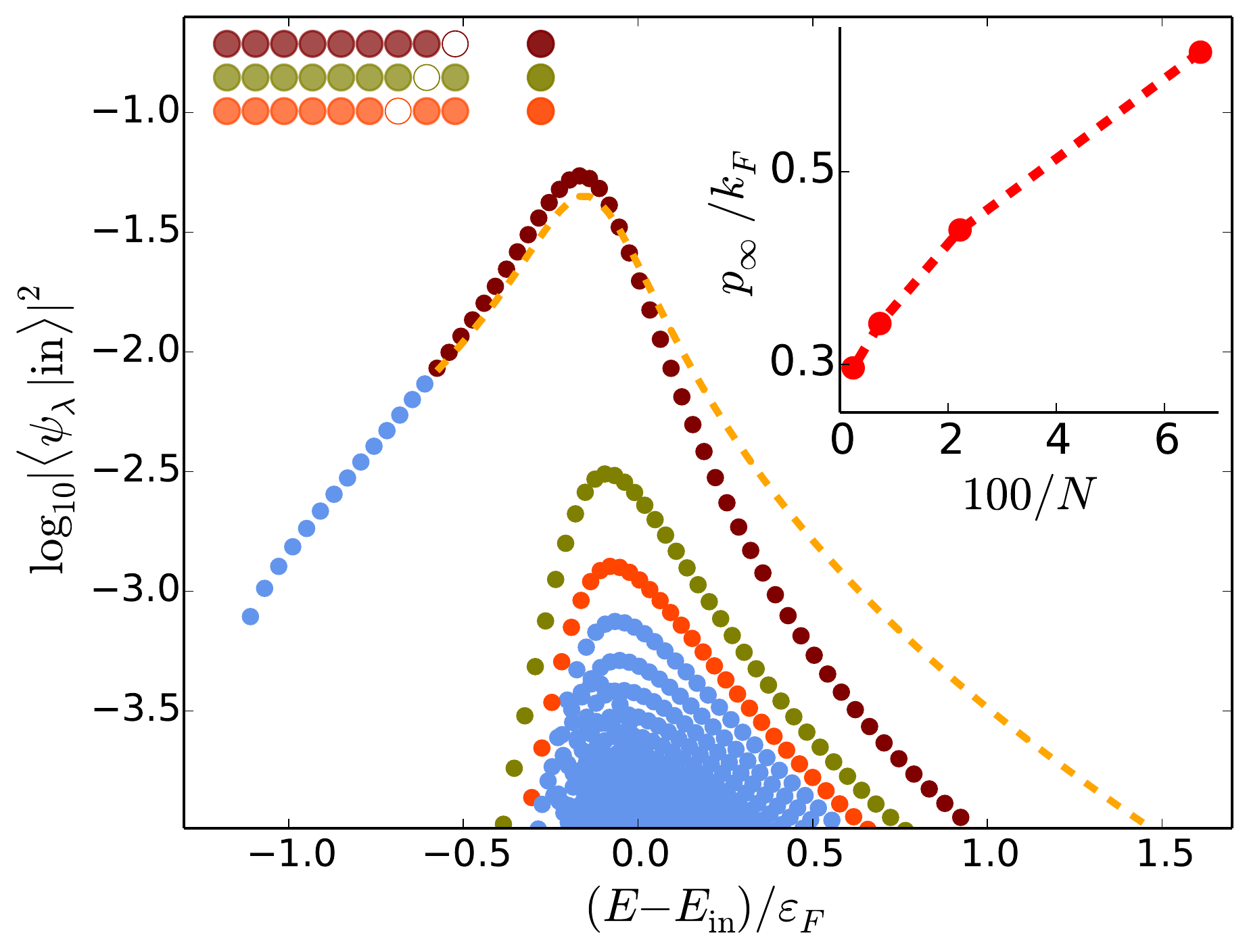}
\caption{(color online).
Overlaps $\left|\la \psi_\lambda |\rm in \ra\right|^2$ versus energy $E$
(relative to the in-state energy $E_{\rm in}$) for $N=135$, $L=405$,
$\gamma=3$, and $p_0 = 1.2 \kF$.
Several families of states are clearly visible. The diagram in the upper left corner
shows schematically the structure of sets $\lambda$ for the
top three families, including the dominant family $\mathfrak{s}$ (maroon).
The dashed curve corresponds to the asymptotic expression
obtained in the limit  $\gamma^2\ln N\to 0,$ $\gamma^2N \to \infty.$ Inset demonstrates the finite-size dependence of~$p_\infty$ at  $p_0 = 1.6 \kF$ and $\gamma=3$.
}
\label{fig:overlaps}
\end{figure}

We do numerics on systems with up to $405+1$ particles, which is an order of
magnitude improvement compared to earlier approaches
\cite{mathy2012quantum,knap2013quantum}. We find a substantial dependence of
$p_\infty$ on $N$, which persists up to the largest available $N$, see Fig. \ref{fig:overlaps}. Thus, a thorough
investigation of finite size corrections is essential for extracting the
thermodynamic limit behavior. We reserve such analysis for a separate
publication \cite{Burovski2013}.

Concentrating on the regime of moderate $\gamma$, we find that  Eq.\
\eqref{ffactors} is dominated by the one-parameter family of states
$\mathfrak{s} = \{\lambda(\widetilde{n}),\;\widetilde{n}> N/2\}$, where $\lambda(\widetilde{n})\equiv	{\scriptsize \{-(N-1)/2,-(N-3)/2,\dots,(N-1)/2,\widetilde{n}\}}$, see Fig.\ \ref{fig:overlaps} for an
illustrative example.
In fact, in the limit $\gamma\to 0$ the family $\mathfrak{s}$  can be investigated analytically.

{\em Bethe Ansatz, asymptotic analysis.}---
In the limit
$\gamma^2 \ln N\to 0$
and
$\gamma^2 N \rightarrow \infty$
we are able to obtain an explicit asymptotic expressions for the form-factors $\la \psi_\lambda | \rm in\ra$ and $\la \psi_\lambda |\hat{P}_i| \rm \psi_\lambda\ra$ for $ \lambda\in\mathfrak{s}$. Furthermore, we prove that the states from $\mathfrak{s}$ saturate
the sum rule,
$\sum_{\lambda\in\mathfrak{s}}| \la \psi_\lambda | \rm in\ra |^2 = 1$. This way, Eq. (\ref{ffactors}) yields \cite{Burovski2013}
\be\label{asymptotic momentum equal masses}
p_\infty=p_0-\theta\left(|p_0|-\kF\right)\frac{p_0^2-\kF^2}{2\kF} \ln \frac{p_0+\kF}{p_0-\kF}\,.
\ee

{\em Discussion and outlook.}---
It is interesting now to compare Eq. (\ref{asymptotic momentum equal masses}) with an exact solution of Eq. \eqref{integral eq for asymptotic momentum}
in the case of equal masses:
\begin{equation}
 p_{\infty}^{B} = p_0 - 2\kF\,\theta\left(|p_0|-\kF\right)  \left( \ln \frac{p_0+\kF}{p_0-\kF} \right)^{-1}\,.
 \label{Bolt}
\end{equation}
We see that the Boltzmann theory fails to produce a correct result at $m_i=m_h$, which is the consequence of the resonant interaction discussed earlier.
In the vicinity of $\gamma=0$, $\eta=1$ point in the $(\gamma,\eta)$ plane the validity of Eqs.~\eqref{asymptotic momentum equal masses} and \eqref{Bolt} depends on the $|\eta-1|/\gamma$ ratio.
Eq. \eqref{asymptotic momentum equal masses} is valid for $|\eta-1|/\gamma \ll 1$ and Eq. \eqref{Bolt} for $|\eta-1|/\gamma \gg 1$. From mathematical point of view this means that the limits $\eta \to 1$ and $\gamma\to 0$ do not commute.  At any finite $\gamma$ there is no discontinuity of $p_{\infty}(\eta)$ at $\eta=1$. This is consistent with findings \cite{mathy2012quantum,knap2013quantum}.

It is interesting to discuss our results in the context of thermalization \cite{Polkovnikov2011}.
When $|p_0|>q_0$ the impurity is kinematically allowed to exchange energy and momentum
with the bath (host). If such exchange had led to a complete thermalization, equipartition would have implied $p_{\infty}=0$. We see, however,
that this is not the case no matter whether or not the model is integrable. This seems to be one of the rare examples of the thermalization failure
in a local, nonintegrable model without disorder (see, {\it e.g.,} the discussion in \cite{Gogolin}).

Finally, we outline directions for further development. Eq.~\eqref{Boltzmann equation} can be generalized to describe the motion of the impurity under an external force and at non-zero temperature.  This way, one can investigate the asymptotical momentum as a function of force and describe the quasi-Bloch oscillations \cite{Gangardt2009} in TG gas at an arbitrary ratio $m_i/m_h\neq 1$ \cite{Gamayun2013}. The case of equal masses, where the Boltzmann equation fails, requires a special treatment. To this end, we developed Bethe Ansatz based tools, which can be extended for studying integrable system with applied force and/or finite couplings.


\acknowledgements{We thank M.~Zvonarev and L.~Glazman for illuminating discussions.
The present work was supported by the ERC grant 279738-NEDFOQ. EB acknowledges
partial support from Lancaster University via ECSG grant SGS/18/01.
OL  acknowledges the partial support via grants RFBR-11-02-00441 and RFBR-12-02-00193,
the grant for Leading Scientific Schools N$^{\circ} $3172.2012.2,
and the Center for Science and Education grant N$^{\circ}$ 8411.
}


\nocite{*}

\bibliography{1D}

\end{document}